%% file: bare_jrnl.tex
\documentclass[sigconf, nonacm]{acmart}

\newcommand{\design}{ExaGEMM}
\newcommand{\teasera}{99.2\%}
\newcommand{\teaserb}{13.29×}
\newcommand{\teaserc}{1.92--13.29×}

\usepackage{hyperref}
\usepackage{amsmath,amsfonts}
\usepackage{algorithm,algorithmic}
\usepackage{listings}
\usepackage{array}
\usepackage{graphicx}
\usepackage{textcomp}
\usepackage{xcolor}
\usepackage{orcidlink}
\usepackage{svg}
\usepackage{enumitem}
\usepackage{subfigure}
    
\usepackage{multirow}
\usepackage{makecell}
\usepackage{booktabs}

\usepackage{siunitx}
\usepackage[font=footnotesize]{caption}
\usepackage{verbatim}

\begin{document}

\title{\design: Exploration Framework for CPU-Driven ML Inference via Associative In-Register Computing for Low-Bit GEMM}

\author{Hyunwoo Oh, Suyeon Jang, Hanning Chen, Sanggeon Yun, Ryozo Masukawa, and Mohsen Imani}
\affiliation{\institution{University of California, Irvine}
\country{}}
\email{{hyunwooo, m.imani}@uci.edu}

\begin{abstract}
Low-bit GEMM is increasingly central to efficient ML inference, yet very-low-bit execution remains a poor fit for conventional CPUs. Practical deployment spans fragmented regimes—from 1/2/4-bit weights to varying activation precision—whose feasibility, reuse opportunity, and support cost differ under fixed SIMD and register-file budgets, making lightweight CPU support selection a first-class design problem. We present \textbf{\design{}}, a workload-aware co-design and exploration framework for CPU-native low-bit GEMM via register-resident LUT execution. The key insight is that existing SIMD datapaths already cover table generation and accumulation; the only new hardware is an in-register select/feed mechanism with explicitly modeled cost. \design{} co-explores parameterized kernels and lightweight SIMD ISA support using analytical models of register feasibility, compute cost, memory traffic, and hardware overhead, pruning the candidate space by \textbf{\teasera} before simulation. It then identifies non-dominated support points and generates ISA specs, gem5 patches, and GEMM kernels for validation. Across representative ML models and CPU targets, \design{} improves latency by \textbf{\teaserb} over software-only baselines, while showing that workload-aware frontier selection is especially important for mixed-precision LLM workloads.
\end{abstract}
\maketitle

\input{Chapters/1_Introduction}
\input{Chapters/2_Backgrounds}
\input{Chapters/3_Design_Space}
\input{Chapters/5_Evaluation}
\input{Chapters/6_Conclusion}

\bibliographystyle{ACM-Reference-Format}
\bibliography{References/intro, References/rw_hw_arch, References/technical_contents}

\end{document}

%% file: Chapters/1_Introduction.tex
\section{Introduction}

Machine learning (ML) inference is increasingly shaped by low-precision matrix multiplication. Across language, vision, and multimodal workloads, quantization reduces model size~\cite{survey_quant, quant_integer}, memory traffic~\cite{awq, atom}, and arithmetic cost, making low-bit GEMM a central deployment kernel. Once weights are pushed to 4-, 2-, or 1-bit precision, efficient execution depends on how low-bit symbols are packed, indexed, accumulated, and mapped onto the target processor---a problem compounded by growing demand for private, offline, and cost-sensitive on-device inference~\cite{survey_edge_ai, survey_edge_llm}. For many such systems the target is a CPU~\cite{nomad, tsar}: separate GPUs, NPUs, or FPGAs introduce area, power, and integration costs that CPU-first or tightly integrated deployments cannot absorb~\cite{onnx}.

Very-low-bit GEMM is, however, a poor fit for conventional CPU execution. Table-driven SIMD methods reduce arithmetic cost by replacing explicit low-bit MACs with a compact activation-derived lookup~\cite{tmac}, but shift the bottleneck to table construction and repeated movement through the memory hierarchy~\cite{veclut}. A deeper difficulty is that low-bit deployment is no longer concentrated around a single operating point~\cite{survey_layerwise, veclut}: practical inference spans W1A8~\cite{w1a8}, W2A8~\cite{bitnet, w2a8}, W4A8~\cite{w4a8}, W4A16~\cite{w4a16}, and mixed layer-wise configurations~\cite{survey_layerwise1, amq, lieq, layerwise0}, each imposing different pressures on table size, accumulation structure, register usage, and implementation cost. Under a fixed SIMD and register-file budget, support that is feasible and effective for one regime can become wasteful, over-specialized, or infeasible in another. A single fixed kernel or narrowly tuned ISA extension struggles to cover this fragmented space, and selecting \emph{which} CPU support points to design and implement has not been treated as a first-class design problem.

Existing approaches leave this support-selection problem unaddressed in complementary ways. Software SIMD kernel methods such as T-MAC~\cite{tmac} and Vec-LUT~\cite{veclut} execute on stock CPUs without ISA changes and target fixed precisions, so the hardware support itself is not a design variable. SIMD ISA-extension methods such as T-SAR~\cite{tsar} do propose microarchitectural support, but commit to a single regime—ternary LLM—leaving broader 1/2/4-bit and mixed-precision ML workloads unsupported. Neither category answers the question a hardware designer faces before tapeout: \textit{across a fragmented workload spanning multiple quantization regimes and implementation-cost constraints, \textbf{which support points are worth implementing in vector CPU designs, and when is one aggressive fixed design no longer enough?}}

We present \textbf{\design{}}, a workload-/budget-aware RTL/ISA/kernel co-design and exploration framework for register-resident LUT-based low-bit GEMM on CPUs. The key insight is that table-driven GEMM can largely reuse existing SIMD addition, dot-product, and accumulation datapaths~\cite{neonrvv, neon_dotp}; the main new hardware is an in-register select mechanism that applies lookup tables to staged weight blocks. Because its cost varies with precision, packing granularity, fan-in, tile shape, and mapping style, support cost becomes an explicit and rankable co-design objective.

\design{} analytically co-explores parameterized GEMM kernels and lightweight ISA support using models of register feasibility, instruction count, memory traffic, compute cost, and hardware overhead. This exploration prunes the candidate support space by \textbf{\teasera} prior to gem5-based evaluation, since detailed cache- and system-level effects are too costly to resolve across the full search space. \design{} then identifies feasible non-dominated support points and materializes them as RTL datapath designs, ISA specifications, gem5 ISA-model patches, and GEMM kernels for system-level validation. In evaluation across representative vision transformer (ViT) and LLM models and vector CPU targets, \design{} improves latency by \textbf{\teaserc} over software-only SIMD baselines, and shows that workload-aware frontier selection is especially important for mixed-precision LLM workloads.

The contributions of this paper are as follows:
\begin{itemize}[leftmargin=*,nosep]
\item We formulate low-bit CPU GEMM acceleration as a \emph{support-selection} problem and propose a parameterized execution space that unifies mapping choices across multiple weight and activation precisions, covering both uniform quantization and inter-layer mixed-precision models.

\item We define a lightweight SIMD-slice support abstraction in which the key new hardware is an in-register select mechanism whose cost scales with precision, packing, fan-in, and tile shape, making hardware overhead an explicit and rankable co-design objective alongside compute cost and memory traffic.

\item We present an end-to-end analytical exploration flow that prunes the candidate support space, identifies the non-dominated support frontier, and generates ISA specifications, gem5 ISA-model patches, and GEMM kernels, validated through microarchitectural simulation across CPU targets based on x86 and ARM ISAs.
\end{itemize}

%% file: Chapters/2_Backgrounds.tex
\section{Background and Motivation}
Consider a dot product between low-bit weights and activations. In a table-driven block of size $e$, the $e$ weights are encoded as a symbol $\gamma \in [0,2^{qe})$, while the current activation group generates a LUT
\begin{equation}
T_{\mathbf{a}}[\gamma] = \sum_{j=0}^{e-1} \mathrm{dec}_q(\gamma,j)\, a_j,
\end{equation}
where $\mathrm{dec}_q(\gamma,j)$ is the $j$-th $q$-bit weight encoded in $\gamma$. Instead of issuing $e$ explicit low-bit MACs, the block contribution is produced by lookup and accumulation. LUT-based low-bit execution trades repeated low-bit arithmetic for a compact activation-derived table.

\begin{figure}[t]
\centering
\includegraphics[width=\linewidth]{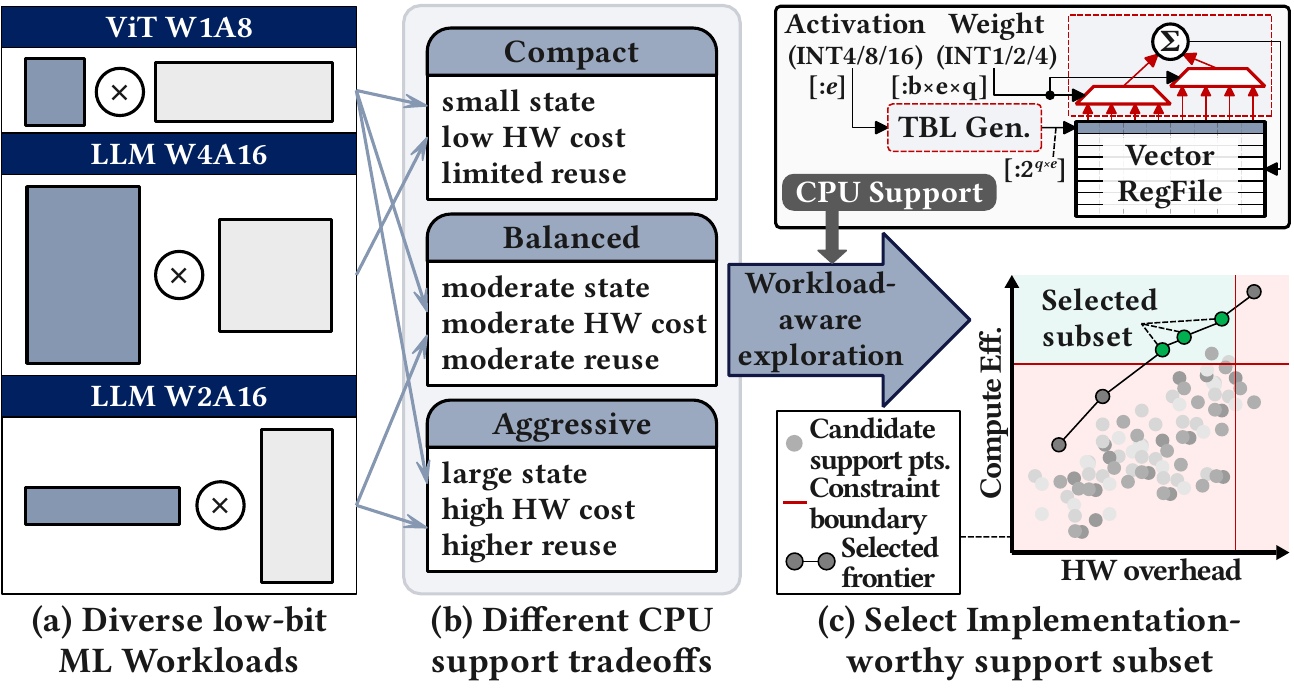}
\caption{Background and motivation for \design{}. Top right block: lightweight CPU support for table-driven low-bit GEMM, where current activation groups generate in-register LUT state that is applied to staged low-bit weight blocks while reusing the SIMD datapath. (a) Diverse low-bit workloads and quantization regimes impose different pressures on table size, reuse, and register-resident state. (b) These pressures lead to different CPU-support tradeoffs, ranging from compact to more aggressive support with higher reuse and higher hardware cost. (c) \design{} therefore performs workload-aware support exploration and selects an implementation-worthy subset of lightweight CPU support points rather than relying on one fixed design point.}
\label{fig:motivation}
\end{figure}

As sketched by the inset of Fig.~\ref{fig:motivation}, this execution style can remain CPU-native. Table generation and application reuse the existing SIMD datapath, while lightweight additional support keeps LUT state in-register and applies it to staged weight blocks. This makes table-driven execution attractive for CPUs, but also introduces tradeoffs absent in conventional dense GEMM.

The first tradeoff is table size. Increasing $e$ improves arithmetic compression, but also grows the table as $2^{qe}$. The second is reuse: reusing a LUT across more rows, columns, or inner products amortizes table-generation cost, but increases register-resident state, aggregation overhead, and pressure on intermediate-result movement. The third is implementation cost: more aggressive support can improve compute efficiency, but requires more hardware state and more specialized logic. Low-bit execution is not merely a lower-precision version of standard SIMD MAC; it is a mapping problem that trades off compression, reuse, state, and hardware cost.

\autoref{fig:motivation} shows why these tradeoffs become a support-selection problem. Practical deployment spans multiple quantization regimes and model mixes, imposing different pressures on reuse, state, and implementation cost. Consequently, no single support point is uniformly best across changing bitwidths, activation precision, and target resource budgets. \autoref{fig:motivation}(b) summarizes this through compact, balanced, and aggressive support classes, while \autoref{fig:motivation}(c) emphasizes the design goal of selecting an implementation-worthy subset from a larger constrained design space.

Prior CPU-oriented methods demonstrate useful operating points \cite{llama_cpp, tmac, nomad, tsar}, but mostly as regime-specific solutions. They show that table-driven or register-resident low-bit execution can be effective on CPUs, yet they do not answer the broader design question raised by the figure: when the target workload spans multiple low-bit GEMM regimes under explicit implementation-cost constraints, which lightweight CPU support points should actually be implemented? As highlighted in Fig.~\ref{fig:motivation}, the candidate support space is too large for exhaustive system-level evaluation, making support selection itself a first-class design problem.

This observation motivates the formulation used in \design{}. Let $\Omega$ denote the candidate space of lightweight CPU support points together with their associated kernel mappings. Rather than searching for one best low-bit kernel, \design{} evaluates feasible points $x \in \Omega$ under workload and hardware constraints, prunes the space analytically, and identifies an implementation-worthy non-dominated subset that balances compute efficiency against hardware overhead, as illustrated by the rightmost panel of Fig.~\ref{fig:motivation}. The next section makes this concrete by defining the parameterized execution space and analytical models for workload-aware support exploration and selection.

%%%%%%%%%%%%%%%%%%%%%%%%%%%%%%%%%%%%%%%%
%%%%%%%%%%%%%%%%%%%%%%%%%%%%%%%%%%%%%%%%

%% file: Chapters/3_Design_Space.tex
%%%%%%%%%%%%%%%%%%%%%%%%%%%%%%%%%%%%%%%%
\begin{figure}[t]
\centering
\includegraphics[width=0.9\linewidth]{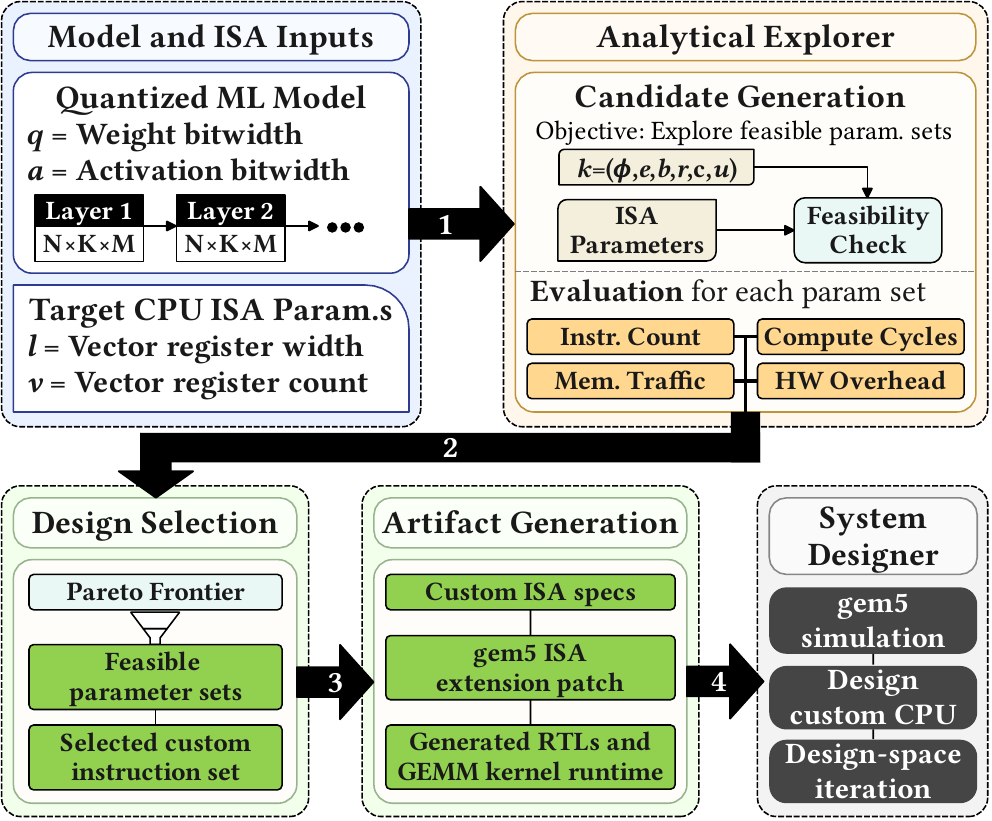}
\caption{
Framework overview of \design{}. Given a quantized ML model and target CPU parameters, \design{} explores feasible parameterized kernels $k=(e,b,r,c,u)$ and evaluates them analytically in terms of instruction count, compute cycles, memory traffic, and hardware overhead. The framework then identifies feasible custom instruction support points, reports the non-dominated support frontier, generates implementation artifacts including ISA specifications, gem5 ISA model patches, and GEMM kernels, and enables designer-guided system-level design-space iteration.
}
\label{fig:framework}
\end{figure}
%%%%%%%%%%%%%%%%%%%%%%%%%%%%%%%%%%%%%%%%

\section{Lightweight ISA Support and Analytical Co-Exploration}
\label{sec:hw_mapping}

\autoref{fig:framework} summarizes the overall \design{} flow.
Given the lightweight SIMD support point and parameterized execution space, \design{} uses an analytical surrogate to estimate register feasibility, instruction count, memory traffic, compute cost, and hardware overhead for each candidate kernel. These estimates are used to prune infeasible candidates, select layer-wise kernels, and report a non-dominated support frontier before system-level validation.

%%%%%%%%%%%%%%%%%%%%%%%%%%%%%%%%%%%%%%%%
\begin{figure*}[t]
\centering
\includegraphics[width=\linewidth]{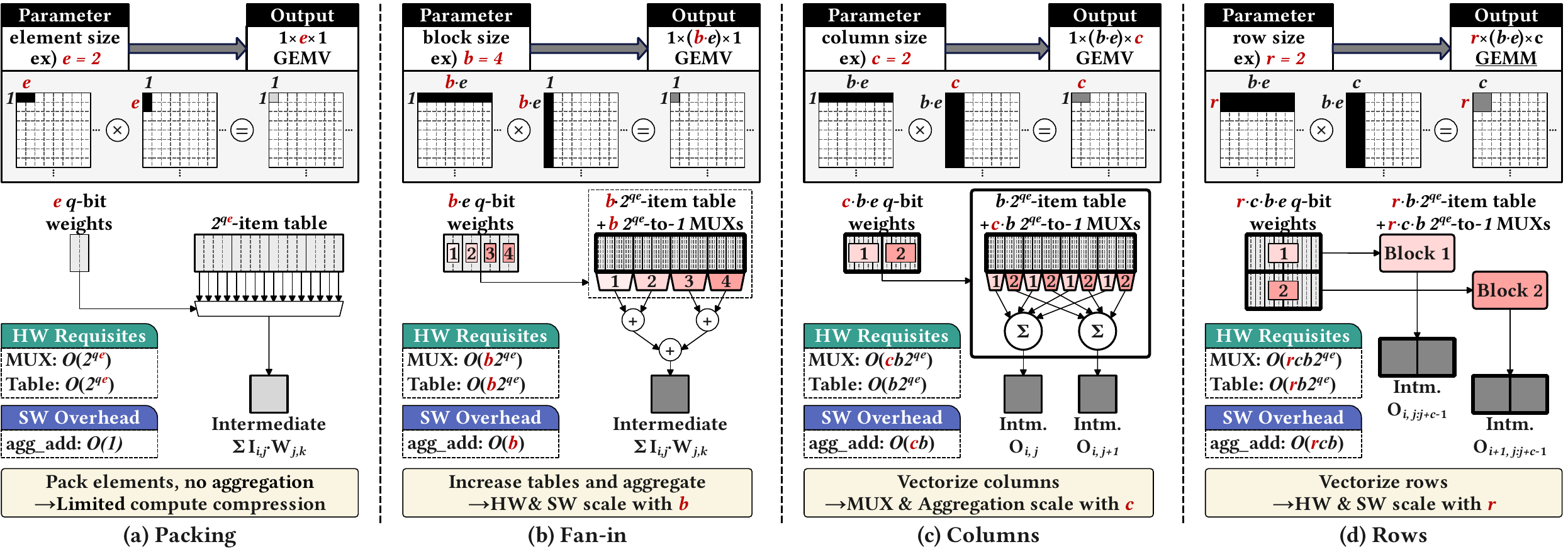}
\caption{
Design-space progression for in-register table-driven computation from GEMV to GEMM. Panels (a)–(d) expand the output dimensionality under quantization bitwidth $q$, by varying \textbf{(a)} packing $e$, \textbf{(b)} fan-in $b$, \textbf{(c)} column tile $c$, and \textbf{(d)} row tile $r$. Each panel reports how MUX/select logic and register-resident table entries scale, along with the extra aggregation-add software overhead (agg\_add) introduced by the decomposition.
}
\label{fig:design_space}
\end{figure*}
%%%%%%%%%%%%%%%%%%%%%%%%%%%%%%%%%%%%%%%%

%%%%%%%%%%%%%%%%%%%%%%%%%%%%%%%%%%%%%%%%%%%%%%%%%%%%%%%%%%%%%%%%%%%%%%%%%%%%%%%%%%%%%%%%%%%%%%%%%%%%%%%%%%%%%%%%%%%%%%%%%%%%%%%
\subsection{ISA Support and Execution Space}
\label{subsec:exec_space}

\design{} is built on two ideas: lightweight ISA support and parameterized execution. Low-bit table-driven GEMM can be supported on vector CPUs by reusing most of the existing SIMD datapath and adding only minimal support for in-register table computation. At the same time, this execution style should not be tied to a single kernel organization. Instead, \design{} exposes a parameterized execution space so that different quantized models and CPU targets can select appropriate operating points. We next describe the framework flow, structural execution space, SIMD-slice support, and software-visible primitives used by the generated kernels.

%%%%%%%%%%%%%%%%%%%%%%%%%%%%%%%%%%%%%%%%%%%%%%%%%%%%%%%%%%%%%%%%%%%%%%%%%%%%%
\subsubsection{\textbf{Framework Overview}}
\label{subsubsec:framework_overview}

\autoref{fig:framework} illustrates the model-\linebreak specific CPU support generation flow in \design{}.
The framework takes as input a quantized ML model, including per-layer bitwidth and GEMM dimensions, together with target CPU parameters such as vector-register width and register-file capacity. With per-layer modeling, the same flow applies to both uniform and inter-layer mixed precision models.
The analytical explorer then enumerates candidate execution instances of the form
\begin{equation}
k=(e,b,r,c,u),
\label{eq:candidate_kernel}
\end{equation}
% where $(e,b,r,c)$ define the structural organization of table-driven GEMM, and $u$ controls instruction-level merging (ILM).
where $(e,b,r,c)$ specify the structural mapping of the table-driven GEMM—packing factor, fan-in, row tile, and column tile, respectively—and u denotes the instruction-level merge factor.
The analytical explorer evaluates candidate kernels $k=(e,b,r,c,u)$ using estimates of feasibility, instruction count, memory traffic, compute cost, and hardware overhead. It then uses these estimates under candidate support points to select layer-wise kernels, identify feasible support points, and report the resulting support frontier to the designer.
These points are materialized as implementation artifacts, including custom ISA specs, gem5 ISA patches, and generated GEMM kernels, for subsequent system-level design-space iteration.

%%%%%%%%%%%%%%%%%%%%%%%%%%%%%%%%%%%%%%%%%%%%%%%%%%%%%%%%%%%%%%%%%%%%%%%%%%%%%
\subsubsection{\textbf{Parameterized Execution Space}}
\label{subsubsec:design_space}

\autoref{fig:design_space} summarizes the structural execution space exposed by \design{}.
The \emph{packing factor} $e$ determines how many low-bit symbols are packed into one table index, the \emph{fan-in} $b$ determines how many packed groups are aggregated within one table-driven block, the \emph{column tile} $c$ determines how many output columns are produced per block, and the \emph{row tile} $r$ determines how many output rows are generated simultaneously.
Together, these parameters control the size of the register-resident table, the complexity of the select network, and the amount of aggregation work required after table application.

The progression in \autoref{fig:design_space} makes the main structural tradeoff explicit.
Increasing $e$ improves packing efficiency but enlarges table cardinality through the $2^{qe}$ term.
Increasing $b$ improves reuse across packed groups but raises aggregation overhead.
Increasing $c$ and $r$ extends reuse across output columns and rows, but also increases the amount of register-resident table state and the complexity of the select network.
Accordingly, no single structural mapping dominates across all bitwidths, layer shapes, and CPU resource envelopes.
Compact mappings reduce aggregation and register pressure, whereas larger GEMM layers can better amortize aggressive tiling.

\autoref{fig:design_space} captures only the \emph{structural} part of the execution space.
Beyond $(e,b,r,c)$, \design{} also includes the ILM factor $u$, a \emph{schedule-level} parameter.
While $(e,b,r,c)$ determine how the table-driven operator is blocked across the GEMM dimensions, $u$ determines how many K-subblocks are merged into one scheduled block.
The $(e,b,r,c)$ define the structural shape of the kernel, while $u$ controls software-visible instruction grouping and register reuse.

%%%%%%%%%%%%%%%%%%%%%%%%%%%%%%%%%%%%%%%%%%%%%%%%%%%%%%%%%%%%%%%%%%%%%%%%%%%%%
\subsubsection{\textbf{Hardware Mapping to a SIMD Slice}}
\label{subsubsec:hw_mapping}

%%%%%%%%%%%%%%%%%%%%%%%%%%%%%%%%%%%%%%%%
\begin{figure}[t]
\centering
\includegraphics[width=0.9\linewidth]{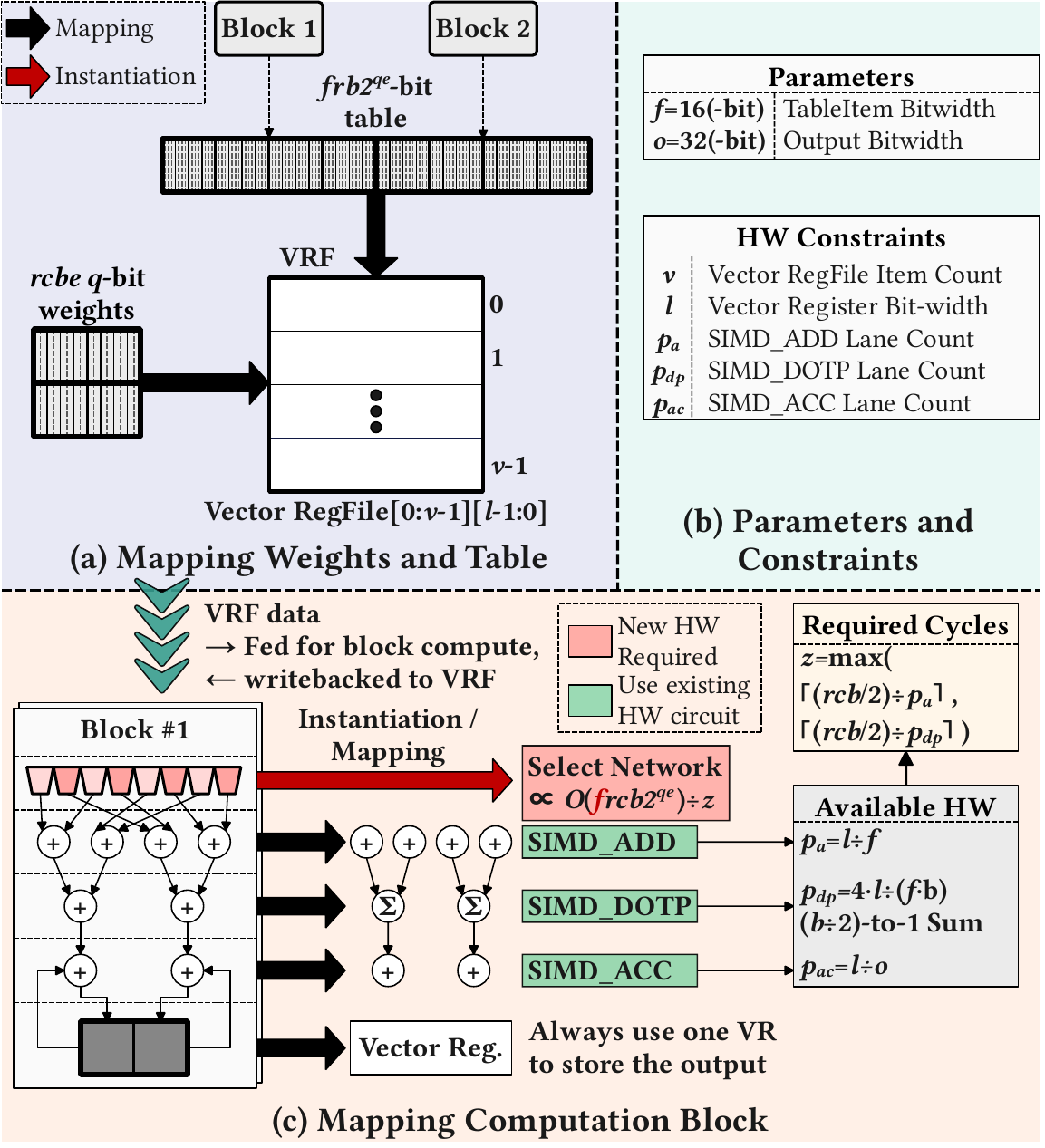}
\caption{
Mapping \design{}’s table-driven block computation onto a vector-CPU SIMD slice. (a) Quantized weights and the $rb2^{qe}$-entry table are staged in the vector register file (VRF). (b) Parameters and hardware constraints define the available lane parallelism. (c) The computation block is realized by instantiating a select network while reusing existing SIMD ADD/DOTP/ACC units; intermediate results are reduced and written back to the VRF.
}
\label{fig:hw_mapping}
\end{figure}
%%%%%%%%%%%%%%%%%%%%%%%%%%%%%%%%%%%%%%%%

Given the structural execution space above, \design{} adopts a lightweight SIMD-slice support point shown in \autoref{fig:hw_mapping}.
The key idea is to realize table-driven low-bit GEMM using \emph{register-resident} data structures and existing SIMD arithmetic datapaths, while introducing only a lightweight select network to perform table application.
This choice is motivated by the desire to keep the required hardware support CPU-native and practical for edge-oriented deployments.

As shown in \autoref{fig:hw_mapping}(a), the vector register file (VRF) stores the live state of execution, including the staged weight block, the register-resident table entries, and the output register.
The table is generated and consumed entirely within the SIMD slice, so table application no longer depends on repeated accesses to memory-resident LUTs.
\autoref{fig:hw_mapping}(b) summarizes the target hardware envelope in terms of vector-register (VR) count, VR width, and the available SIMD lane counts for the reused arithmetic units.
These parameters define the constraints under which candidate kernels must operate.

\autoref{fig:hw_mapping}(c) then shows the key architectural property of \design{}:
The major new hardware support is a \emph{select network}, while the existing SIMD\_ADD, SIMD\_DOTP, and SIMD\_ACC circuits are reused for table generation, table-driven compute, and accumulation, respectively.
If the required computation for each instruction exceeds the existing circuits, a multi-cycle counter controls the data feeding.
This mapping keeps the support point lightweight and makes hardware overhead explicit: the select network scales with the amount of table-driven indirection that a candidate kernel requires, while the remaining computation is serviced by the existing SIMD execution pipeline.
In other words, \design{} does not propose a new accelerator array; it exposes a CPU-native support point that extends the SIMD slice just enough to make in-register table-driven low-bit GEMM practical.

This hardware organization naturally induces two execution primitives in \design{}.
The VRF-resident table is generated from packed activation state and then consumed together with a weight block to update an output tile.
\autoref{tab:logical_primitives} summarizes the resulting table-generation and table-driven GEMM primitives.

%%%%%%%%%%%%%%%%%%%%%%%%%%%%%%%%%%%%%%%%%%%%%%%%%%%%%%%%%%%%%%%%%%%%%%%%%%%%%
\subsubsection{\textbf{Instruction Primitives and Instruction-Level Merging}}
\label{subsubsec:instr_merge}

The logical primitives in \autoref{tab:logical_primitives} provide the interface between the lightweight SIMD ISA extension and the generated GEMM/GEMV kernels.
In the baseline case, one \texttt{TBL\_r$\times$be} instruction generates a table for one K-subblock, and one or more \texttt{GEMM\_r$\times$be$\times$c} instances use that table to update the corresponding output tiles.
However, executing these primitives naively can introduce redundant output load/store traffic and loop overhead, especially when multiple neighboring K-subblocks contribute to the same output tile.

%%%%%%%%%%%%%%%%%%%%%%%%%%%%%%%%%%%%%%%%
\begin{table}[t]
\centering
\caption{Logical ISA primitives induced by the lightweight SIMD support point.}
\label{tab:logical_primitives}
\small
\setlength{\tabcolsep}{5pt}
\resizebox{\linewidth}{!}{
\begin{tabular}{l p{7.8cm}}
\toprule
\textbf{Primitive} & \textbf{Logical effect} \\
\midrule
\texttt{TBL\_r$\times$be} &
Consumes a packed activation block and generates an in-register table in the VRF for subsequent table-driven computation. \\

\texttt{GEMM\_r$\times$be$\times$c} &
Consumes one in-register table and one weight block to update an $r\times c$ output tile through the select network and reused SIMD datapaths. \\
\bottomrule
\end{tabular}
}
\end{table}
%%%%%%%%%%%%%%%%%%%%%%%%%%%%%%%%%%%%%%%%

%%%%%%%%%%%%%%%%%%%%%%%%%%%%%%%%%%%%%%%%
\begin{figure}[t]
\centering
\includegraphics[width=\linewidth]{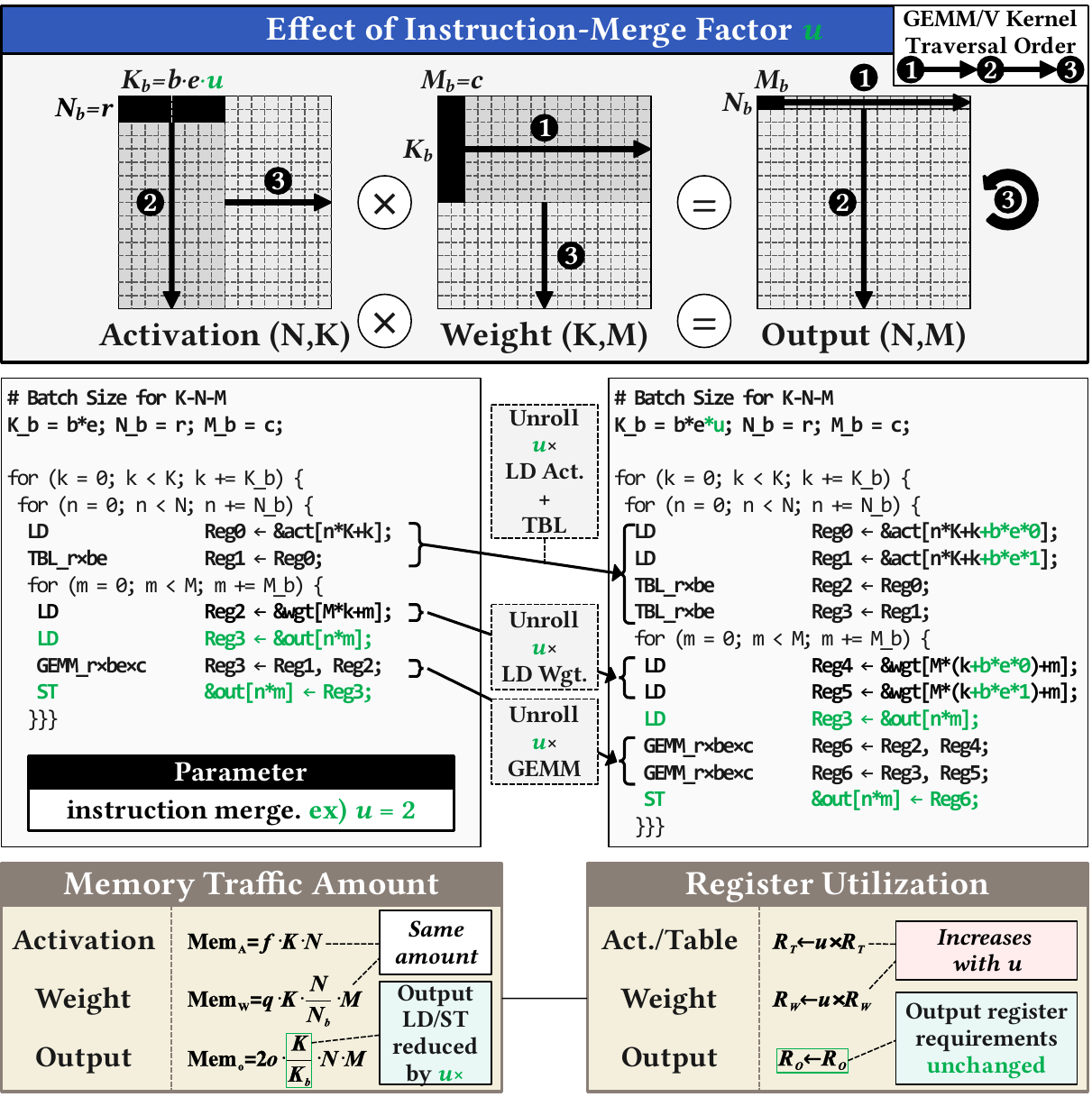}
\caption{
Effect of instruction-level merge (ILM) factor $u$ on scheduled table-driven GEMM execution.
Increasing $u$ merges multiple K-subblocks into one scheduled block, reducing redundant output load/store traffic and loop overhead while increasing live activation/table and weight register demand. The output-register requirement remains unchanged because all merged K-subblocks accumulate into the same output object.
}
\label{fig:instr_merge}
\end{figure}
%%%%%%%%%%%%%%%%%%%%%%%%%%%%%%%%%%%%%%%%

To address this, \design{} introduces an instruction-level merge (ILM) factor $u$ (\autoref{fig:instr_merge}).
Increasing $u$ merges multiple K-subblocks into one scheduled block, so that
\begin{equation}
K_b = b\cdot e \cdot u,\qquad N_b=r,\qquad M_b=c.
\end{equation}
Relative to the baseline case $u\,=\,1$, the merged schedule keeps activation-derived table state live for a longer window, while reusing the same output VR across multiple \texttt{GEMM\_r$\times$be$\times$c} instances.
As shown in \autoref{fig:instr_merge}, this reduces redundant output LD/ST activity and loop overhead while increasing live activation/table and weight register demand.
The output-register requirement remains unchanged because all merged K-subblocks still accumulate into the same output tile.

This distinction between structural and schedule-level parameters is important for \design{}.
The tuple $(e,b,r,c)$ determines the table-driven block shape and associated hardware requirements, while $u$ controls how aggressively the generated kernel amortizes output traffic and loop overhead.
Together, they define the full execution candidate $k=(e,b,r,c,u)$ evaluated by the analytical explorer in the following subsection.

%%%%%%%%%%%%%%%%%%%%%%%%%%%%%%%%%%%%%%%%%%%%%%%%%%%%%%%%%%%%%%%%%%%%%%%%%%%%%%%%%%%%%%%%%%%%%%%%%%%%%%%%%%%%%%%%%%%%%%%%%%%%%%%
\subsection{Analytical Exploration Framework}
\label{subsec:exploration}

Given the lightweight SIMD support point and parameterized execution space in \autoref{subsec:exec_space}, \design{} evaluates which candidate kernels are feasible and worth supporting for a target quantized model.
For each candidate, the analytical explorer models register feasibility, logical instruction count, memory traffic, arithmetic/control cost, and hardware overhead.
These quantities are used to prune infeasible candidates, rank feasible kernels, and report a non-dominated support frontier before system-level validation.

%%%%%%%%%%%%%%%%%%%%%%%%%%%%%%%%%%%%%%%%%%%%%%%%%%%%%%%%%%%%%%%%%%%%%%%%%%%%%
\subsubsection{\textbf{Feasibility and Cost Models}}
\label{subsubsec:analytic_model}

For a quantized GEMM layer $\ell$ of shape $(N_\ell,K_\ell,M_\ell)$ and bitwidth $q_\ell$, \design{} evaluates each candidate
$k=(e,b,r,c,u)$ in \autoref{eq:candidate_kernel}.
Modeling $q_\ell$ per layer allows the same framework to cover both uniform- and mixed-quantization settings.

\noindent\textbf{Register feasibility.}
For candidate $k$, the modeled VRF-resident live state consists of the register-resident table, weight block, and output tile:
\begin{equation}\label{eq:reg_usage}
\resizebox{0.92\linewidth}{!}{$
R_T(k) = \left\lceil \frac{f \, r \, b \, 2^{q_\ell e} \, u}{l} \right\rceil,\quad
R_W(k) = \left\lceil \frac{q_\ell \, r \, c \, b \, e \, u}{l} \right\rceil,\quad
R_O(k) = \left\lceil \frac{o \, r \, c}{l} \right\rceil,
$}
\end{equation}
where $l$ is the vector-register width, $f$ is the table-entry width, and $o$ is the output / accumulation width.
The factor $u$ appears because ILM increases simultaneously live table and weight state.
We define
\begin{equation}
U_{\mathrm{VRF}}(k)=\frac{R_T(k)+R_W(k)+R_O(k)}{v},
\label{eq:feasibility_metrics}
\end{equation}
and require
\begin{equation}
U_{\mathrm{VRF}}(k)\le 1,\qquad R_O(k)\le 1.
\label{eq:feasibility}
\end{equation}
The first constraint enforces total VRF capacity; the second enforces the single-destination-VR execution model assumed by \design{}.

\noindent\textbf{Instruction and traffic models.}
For compactness, define
\begin{equation}
\nu_\ell(k)=\left\lceil \frac{N_\ell}{r} \right\rceil,
\kappa_\ell(k)=\left\lceil \frac{K_\ell}{be} \right\rceil,
\mu_\ell(k)=\left\lceil \frac{M_\ell}{c} \right\rceil,
\kappa_\ell^{(u)}(k)=\left\lceil \frac{K_\ell}{beu} \right\rceil.
\label{eq:helper_terms}
\end{equation}
The dominant logical instruction counts are
\begin{equation}\label{eq:tbl_gemm_count}
I_{\mathrm{TBL}}(\ell,k)=\nu_\ell(k)\kappa_\ell(k), \qquad
I_{\mathrm{GEMM}}(\ell,k)=\nu_\ell(k)\kappa_\ell(k)\mu_\ell(k).
\end{equation}
ILM changes the schedule rather than the logical work decomposition, so $u$ does not reduce the number of \texttt{TBL} or \texttt{GEMM} primitives.
Instead, it amortizes output load/store activity and loop overhead:
\begin{equation}
I_{\mathrm{LD}}^{o}(\ell,k)=I_{\mathrm{ST}}^{o}(\ell,k)=
\nu_\ell(k)\kappa_\ell^{(u)}(k)\mu_\ell(k),
\label{eq:output_ldst}
\end{equation}
\begin{equation}
C_{\mathrm{CTL}}(\ell,k)=
\alpha_{\mathrm{CTL}}\,
\nu_\ell(k)\kappa_\ell^{(u)}(k)\mu_\ell(k),
\label{eq:control_overhead}
\end{equation}
where $\alpha_{\mathrm{CTL}}$ is the platform-dependent control-overhead cost per scheduled block.

\design{} models schedule-level memory traffic without attempting to model the full cache hierarchy.
At fixed $(e,b,r,c)$, activation and weight traffic are invariant with respect to $u$, while output traffic decreases as output LD/ST is amortized across merged K-subblocks:
\begin{equation}
B_o(\ell,k)=
o\,r\,c\left(I_{\mathrm{LD}}^{o}(\ell,k)+I_{\mathrm{ST}}^{o}(\ell,k)\right),
\label{eq:output_traffic}
\end{equation}
\begin{equation}
B_{\mathrm{mem}}(\ell,k)=B_a(\ell,k)+B_w(\ell,k)+B_o(\ell,k),
\label{eq:mem_total}
\end{equation}
where $B_a$ and $B_w$ denote activation and weight traffic under the selected structural mapping.

\noindent\textbf{Execution cost and hardware overhead.}
Let $\tau_{\mathrm{TBL}}(k)$ and $\tau_{\mathrm{GEMM}}(k)$ denote the per-instruction costs of the two logical primitives.
Because \texttt{GEMM} reuses the SIMD\_ADD, SIMD\_DOTP, and SIMD\_ACC datapaths, its per-instruction cost is bounded by the slowest reused arithmetic component:
\begin{equation}
\tau_{\mathrm{GEMM}}(k)=
\max\!\left(
\tau_{\mathrm{ADD}}(k),
\tau_{\mathrm{DOTP}}(k),
\tau_{\mathrm{ACC}}(k)
\right).
\label{eq:gemm_cost}
\end{equation}
The arithmetic/control-side estimate is
\begin{equation}
\begin{aligned}
C_{\mathrm{cmp}}(\ell,k)=\;&
I_{\mathrm{TBL}}(\ell,k)\tau_{\mathrm{TBL}}(k)
+ I_{\mathrm{GEMM}}(\ell,k)\tau_{\mathrm{GEMM}}(k) \\
&+ I_{\mathrm{agg}}(\ell,k)\tau_{\mathrm{ADD}}(k)
+ C_{\mathrm{CTL}}(\ell,k),
\end{aligned}
\label{eq:compute_cost}
\end{equation}
where $I_{\mathrm{agg}}$ denotes the aggregation-add work by the structural mapping in \autoref{fig:design_space}.
$C_{\mathrm{cmp}}$ captures arithmetic/control work only, while memory-side effects enter the final ranking objective by $B_{\mathrm{mem}}$.

To rank candidates, \design{} combines arithmetic/control cost and schedule-level traffic into a single analytical execution cost:
\begin{equation}
C_{\mathrm{exe}}(\ell,k)=
C_{\mathrm{cmp}}(\ell,k)+\alpha_{\mathrm{mem}}\,B_{\mathrm{mem}}(\ell,k),
\label{eq:efficiency_score}
\end{equation}
where $\alpha_{\mathrm{mem}}$ is a platform-dependent traffic-to-cost factor that maps the schedule-level traffic proxy $B_{\mathrm{mem}}$ to an effective execution-cost penalty.
$\alpha_{\mathrm{mem}}$ is obtained by calibrating the analytical score against vectorized load/store microbenchmarks that sweep working-set sizes across the cache and memory hierarchy.
$C_{\mathrm{exe}}$ is the final execution-side ranking objective used during exploration.
It is not intended to predict the full memory hierarchy exactly; rather, it provides a consistent efficiency ordering before detailed simulation.

Finally, hardware overhead is modeled per supported primitive configuration
$p=(e,b,r,c,q)$ as
\begin{equation}
H_{\mathrm{sel}}(p)=\alpha_{\mathrm{sel}}\,S_{\mathrm{sel}}(e,b,r,c,q),
\label{eq:hw_overhead}
\end{equation}
where $S_{\mathrm{sel}}$ is the select-network scaling term corresponding to \autoref{fig:design_space}, and $\alpha_{\mathrm{sel}}$ is a technology-dependent factor that maps $S_{\mathrm{sel}}$ to normalized selector overhead using ASIC synthesis under the target process design kit.

Taken together, \autoref{eq:feasibility}--\autoref{eq:hw_overhead} define a compact analytical surrogate for model-specific ISA support exploration.
\design{} uses this surrogate both to rank candidate kernels and to identify lightweight support points worth materializing for subsequent gem5-based evaluation.

%%%%%%%%%%%%%%%%%%%%%%%%%%%%%%%%%%%%%%%%%%%%%%%%%%%%%%%%%%%%%%%%%%%%%%%%%%%%%
\subsubsection{\textbf{Kernel Selection and Support Ranking}}
\label{subsubsec:selection}

Using the models above, \design{} performs exploration at two levels: \textit{(i)} layer-wise kernel selection and \textit{(ii)} model-level support-point ranking.

\noindent\textbf{Layer-wise kernel selection.}
For each support point $\sigma$ and layer $\ell$, \design{} enumerates the candidate set $\mathcal{K}_\ell(\sigma)$ over the supported ranges of $(e,b,r,c,u)$.
Candidates violating \autoref{eq:feasibility} are pruned.
Among the remaining candidates, \design{} selects
\begin{equation}
\begin{aligned}
k_\ell^\star(\sigma)
= \arg\min_{k\in\mathcal{K}_\ell(\sigma)}\;& C_{\mathrm{exe}}(\ell,k) \\
\text{s.t.}\;& U_{\mathrm{VRF}}(k)\le 1,\;
R_O(k)\le 1,
\end{aligned}
\label{eq:layer_select}
\end{equation}
where $C_{\mathrm{exe}}$ is defined in \autoref{eq:efficiency_score}.
Hardware overhead is a property of the support point rather than of an individual layer mapping, so layer-wise selection minimizes execution-side cost under feasibility constraints, while the efficiency--hardware tradeoff is resolved at the support-point level.

\noindent\textbf{Support-point frontier.}
A custom ISA support point $\sigma$ is defined as a set of distinct supported primitive configurations $p=(e,b,r,c,q)$.
The ILM factor $u$ is not part of $\sigma$ because it changes the software schedule but does not require a different primitive.
For each $\sigma$, \design{} restricts layer-wise selection to compatible candidates and discards the support point if any layer has no feasible mapping.
Otherwise, the selected per-layer kernels are aggregated into model-level execution cost and hardware overhead:
\begin{equation}
C_{\mathrm{model}}(\sigma)=\sum_{\ell} C_{\mathrm{exe}}(\ell,k_\ell^\star(\sigma)),
\label{eq:model_cost}
\end{equation}
\begin{equation}
H_{\mathrm{model}}(\sigma)=\sum_{p\in\sigma} H_{\mathrm{sel}}(p),
\label{eq:model_hw}
\end{equation}
where $p$ denotes one supported configuration in the support point.
\autoref{eq:model_hw} uses an additive support-cost model; if multiple configurations share hardware, it should be interpreted as a conservative ranking proxy.
For completeness, \design{} may still report
\begin{equation}
B_{\mathrm{model}}(\sigma)=\sum_{\ell} B_{\mathrm{mem}}(\ell,k_\ell^\star(\sigma)),
\label{eq:model_mem_report}
\end{equation}
but $B_{\mathrm{model}}$ is diagnostic only, since traffic is already incorporated into $C_{\mathrm{model}}$ through \autoref{eq:efficiency_score}.

\design{} reports the non-dominated feasible support points over the efficiency--hardware tradeoff, leaving the final choice to the designer.
Shortlisted points are then materialized as ISA specifications, gem5 ISA-model patches, and generated GEMM kernels for system-level validation.
Algorithm~\ref{alg:selection} summarizes the model-level exploration procedure.

\begin{algorithm}[tbh!]
\caption{Model-level support exploration in \design{}}
\label{alg:selection}
\small
\begin{algorithmic}[1]
\REQUIRE Quantized model $\mathcal{L}$, target CPU parameters, support-point set $\Sigma$
\ENSURE Non-dominated feasible support points
\FOR{each $\sigma \in \Sigma$}
    \STATE feasible $\gets$ true
    \FOR{each layer $\ell \in \mathcal{L}$}
        \STATE Enumerate $\mathcal{K}_\ell(\sigma)$
        \STATE Prune candidates violating \autoref{eq:feasibility}
        \IF{$\mathcal{K}_\ell(\sigma)=\varnothing$}
            \STATE feasible $\gets$ false
            \STATE \textbf{break}
        \ENDIF
        \STATE Select $k_\ell^\star(\sigma)$ using \autoref{eq:layer_select}
    \ENDFOR
    \IF{feasible}
        \STATE Aggregate $C_{\mathrm{model}}(\sigma)$ and $H_{\mathrm{model}}(\sigma)$
        \STATE Optionally record $B_{\mathrm{model}}(\sigma)$
    \ENDIF
\ENDFOR
\STATE Extract the non-dominated feasible points over $(C_{\mathrm{model}}, H_{\mathrm{model}})$
\STATE Return the resulting support frontier to the designer
\end{algorithmic}
\end{algorithm}

This procedure reduces the model-specific support space to a small set of feasible, non-dominated candidates for designer-guided selection and subsequent gem5-based evaluation.

%% file: Chapters/5_Evaluation.tex
\vspace{-1mm}
\section{Evaluation}
\label{sec:eval}

We evaluate \design{} from both CAD and system perspectives by asking:
(1) whether the analytical explorer prunes the search space while preserving strong candidates,
(2) whether realistic workloads induce different feasible and execution-favorable primitive-side landscapes under fixed target constraints,
(3) whether frontier-selected support points and kernel schedules improve end-to-end performance under modest hardware overhead.

%%%%%%%%%%%%%%%%%%%%%%%%%%%%%%%%%%%%%%%%%%%%%%%%%%%%%%%%%%%%%%%%%%%%%%%%%%%%%%%%%%%%%%%%%%%%%%%%%%%%%%%%%%%%%%%%%%%%%%%%%%%%%%%
\subsection{Experimental Setup}
\label{subsec:eval_setup}

Table~\ref{tab:system_cfg} and \ref{tab:model_targets} summarize the target platforms and evaluated workloads. We study representative x86 and ARM vector CPU targets on gem5-AVX \cite{gem5, gem5-avx} v20.1.0.0, spanning 128/256/512-bit SIMD configurations with different register budgets. GEMM tiles are distributed across all available cores along the output-row dimension; since each core's in-register table is generated and consumed locally, inter-core coherence overhead is minimized. The workloads span two uniform-quantization cases (DeiT-B W1A8, Llama2-7B W2A16), and one mixed-precision configuration (Llama2-7B with AMQ-style inter-layer W{1,2,4}A16 \cite{amq}), exploring workload-aware support selection across quantization regimes. For each target, \design{} explores candidate kernels $k=(e,b,r,c,u)$, materializes selected frontier points, and evaluates them via model-level support-tradeoff analysis and end-to-end comparison against representative baselines. The analytical surrogate is calibrated per target/process node: $\alpha_{\mathrm{mem}}$ is obtained from gem5 microbenchmarks, and $\alpha_{\mathrm{sel}}$ is derived from ASIC synthesis of the selector-network and full SIMD-datapath RTL designs in a TSMC 28\,nm process at 1\,GHz using Cadence Genus 21.10. Area and power are reported at tt0p9v25c.

\input{Tables/system_cfg}

%%%%%%%%%%%%%%%%%%%%%%%%%%%%%%%%%%%%%%%%
\begin{table}[t]
\centering
\caption{Model targets used in the evaluation.}
\label{tab:model_targets}
\small
\setlength{\tabcolsep}{4pt}
\resizebox{0.75\linewidth}{!}{
\begin{tabular}{l l c l}
\toprule
\textbf{Model} & \textbf{Modality} & \textbf{Label} & \textbf{Quantization setting} \\
\midrule
\textbf{DeiT-B}     & Vision   & \textbf{A} & W1A8 uniform \\
\textbf{Llama2-7B} & Language & \textbf{B} & W2A16 uniform \\
\textbf{Llama2-7B} & Language & \textbf{C} & W\{1,2,4\}A16 mixed \\
\bottomrule
\end{tabular}}
\end{table}
%%%%%%%%%%%%%%%%%%%%%%%%%%%%%%%%%%%%%%%%

%%%%%%%%%%%%%%%%%%%%%%%%%%%%%%%%%%%%%%%%%%%%%%%%%%%%%%%%%%%%%%%%%%%%%%%%%%%%%%%%%%%%%%%%%%%%%%%%%%%%%%%%%%%%%%%%%%%%%%%%%%%%%%%
\subsection{Explorer Effectiveness and Fidelity}
\label{subsec:eval_explorer}

We evaluate the analytical explorer along two axes: (1) how aggressively \design{} reduces the candidate space before simulation, and (2) whether the analytical ranking metric preserves the ordering needed for shortlist formation.

%%%%%%%%%%%%%%%%%%%%%%%%%%%%%%%%%%%%%%%%%%%%%%%%%%%%%%%%%%%%%%%%%%%%%%%%%%%%%
\subsubsection{\textbf{Search Space Reduction}}
\begin{figure}[t]
\centering
\includegraphics[width=1.04\linewidth]{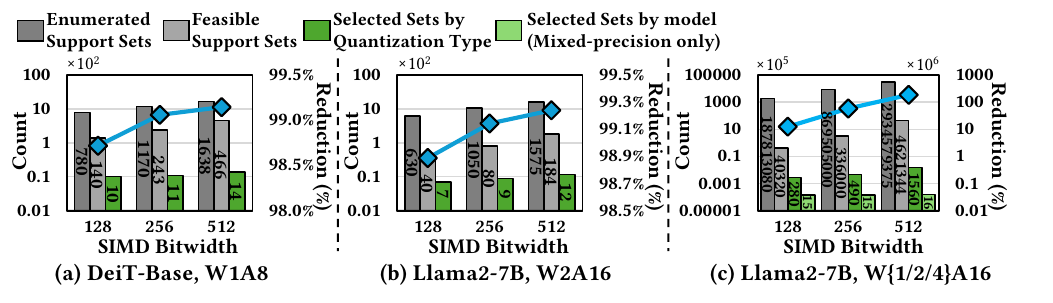}
\caption{Explorer-driven search-space reduction. For DeiT-B W1A8 and Llama2-7B W2A16, bars show enumerated primitives, VRF-feasible primitives, and retained frontier points. For Llama2-7B W\{1/2/4\}A16, bars show the hierarchical mixed-precision flow: enumerated support sets, post-per-quantization candidate sets, and final model-level frontier sets. The line reports overall reduction from the fully enumerated space.}
\label{fig:eval_explorer_reduce}
\end{figure}

\autoref{fig:eval_explorer_reduce} shows that \design{} shrinks large candidate spaces to small frontiers before gem5 runs. For uniform quantization, DeiT-B W1A8 shrinks from 780/1170/1638 enumerated primitive candidates to 10/11/14 frontier points on the 128/256/512-bit targets, corresponding to 98.7\%/99.1\%/99.1\% reduction. Llama2-7B W2A16 similarly shrinks from 630/1050/1575 candidates to 7/9/12 frontier points, i.e., 98.9\%/99.1\%/99.2\% reduction.

\begin{figure}[t]
\centering
\includegraphics[width=0.75\linewidth]{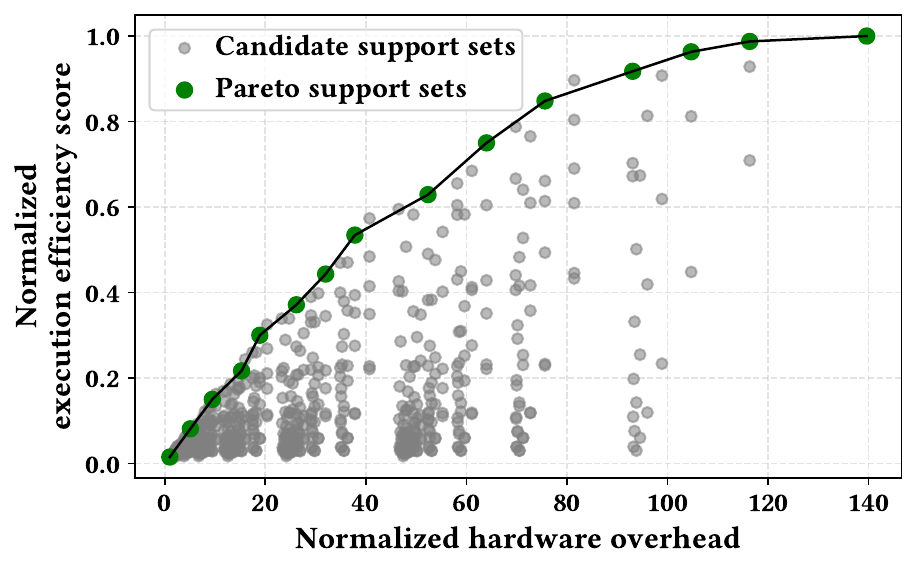}
\caption{Mixed-precision support-set tradeoff space on the 256-bit target. Each point denotes a candidate mixed-precision support set formed by combining Pareto-efficient primitive points from the individual quantization regimes for Llama2-7B W\{1/2/4\}A16. Highlighted points form the final model-level Pareto shortlist retained for materialization and gem5 validation. Execution efficiency is defined as the inverse of execution cost.}
\label{fig:eval_explorer_scatter}
\end{figure}

For mixed precision, \design{} applies the same idea hierarchically. It first computes per-quantization primitive frontiers, forms support-set candidates only from those reduced spaces, and then applies a model-level Pareto filter. For the Llama2-7B W\{1,2,4\}A16, the raw space contains 187{,}813{,}080/869{,}505{,}000/2{,}934{,}579{,}375 structurally enumerated support-set combinations on the 128/256/512-bit targets. After per-quantization frontier reduction, this shrinks to 280/490/1560 candidates, and the final frontier retains only 14--16 sets per target, yielding 12{,}520{,}872$\times$/57{,}967{,}000$\times$/183{,}411{,}211$\times$ reduction before gem5. \autoref{fig:eval_explorer_scatter} presents the 256-bit case: only a small subset of the reduced combinations remains Pareto-competitive.

%%%%%%%%%%%%%%%%%%%%%%%%%%%%%%%%%%%%%%%%
%%%%%%%%%%%%%%%%%%%%%%%%%%%%%%%%%%%%%%%%

%%%%%%%%%%%%%%%%%%%%%%%%%%%%%%%%%%%%%%%%
%%%%%%%%%%%%%%%%%%%%%%%%%%%%%%%%%%%%%%%%

%%%%%%%%%%%%%%%%%%%%%%%%%%%%%%%%%%%%%%%%%%%%%%%%%%%%%%%%%%%%%%%%%%%%%%%%%%%%%
\subsubsection{\textbf{Modeling Fidelity}}

\begin{figure}[t]
\centering
\includegraphics[width=1.065\linewidth]{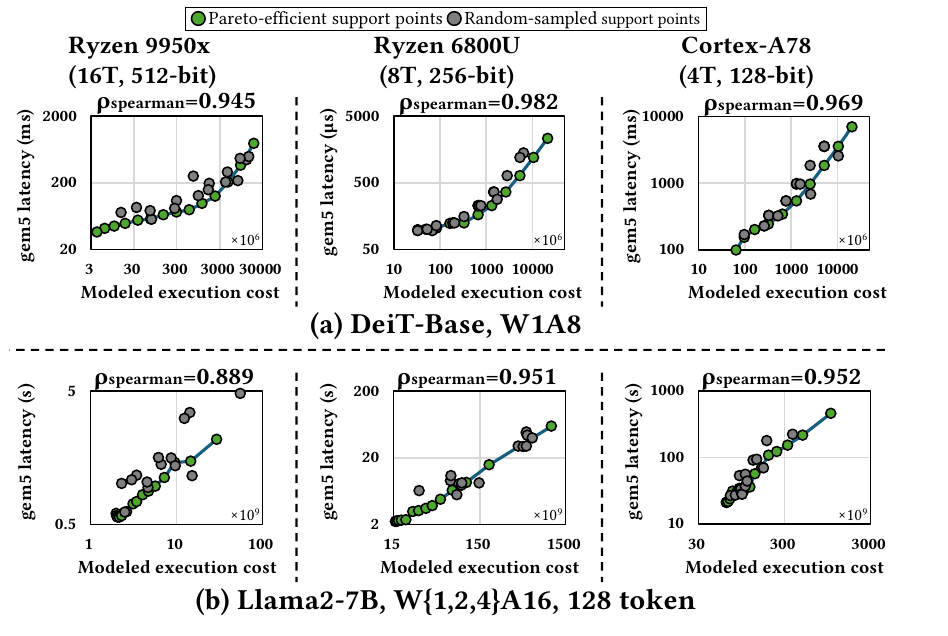}
\caption{Ranking fidelity of the analytical explorer. Each panel compares the modeled execution cost with gem5-measured latency for analytically retained Pareto points and randomly sampled feasible points. Spearman correlations of 0.889--0.982 indicate strong ordering fidelity.}
\label{fig:eval_fidelity_compute}
\end{figure}

We evaluate ranking fidelity as whether the modeled execution cost ($C_{\mathrm{exe}}$) preserves the ordering needed to screen candidates before detailed simulation. For each workload-target pair, we compare the $C_{\mathrm{exe}}$ against gem5 latency on a sample set comprising all analytically retained Pareto points together with randomly sampled feasible points, for a total of 30 points per pair. As shown in \autoref{fig:eval_fidelity_compute}, the Spearman correlations are 0.945/0.982/0.969 for DeiT-B W1A8 and 0.889/0.951/0.952 for mixed-precision Llama2 on the 512/256/128-bit targets.

\begin{figure}[t]
\centering
\includegraphics[width=1.025\linewidth]{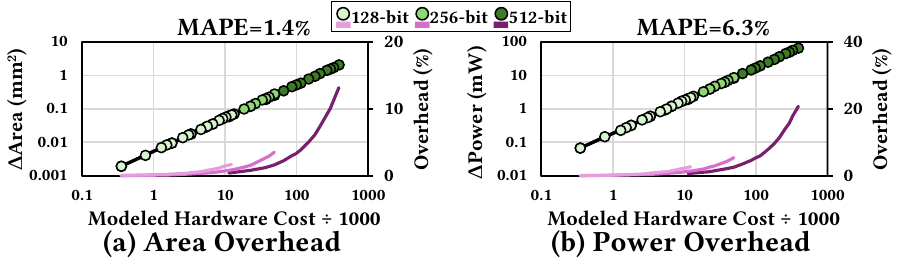}
\caption{Hardware-cost calibration against TSMC 28\,nm synthesis at 1\,GHz for Llama2-7B W\{1,2,4\}A16 support sets on 128/256/512-bit targets. Each point compares the modeled hardware cost with the synthesized area/power overhead for a single support point.}
\label{fig:eval_fidelity_hardware}
\end{figure}

We also calibrate the modeled hardware cost with ASIC synthesis results on the augmented SIMD datapath RTL for the Llama2-7B W\{1,2,4\}A16 Pareto support sets across the 128/256/512-bit targets. \autoref{fig:eval_fidelity_hardware} compares modeled costs with area and power overheads, showing near-linear correlation. The modeled cost, normalized to the smallest synthesized support point's area/power, tracks synthesis closely, with a mean absolute percentage error (MAPE) of 1.4\% for area and 6.3\% for power. Relative to the original full SIMD datapath baseline, retained support points incur 0.049--1.714\% / 0.083--2.731\% area/power overhead on 128-bit, 0.024--3.564\% / 0.042--5.448\% on 256-bit, and 0.381--13.21\% / 0.607--20.81\% on 512-bit, confirming that the added select/feed support remains affordable.

Together, these results make \design{} effective for pre-silicon screening, preserving candidate ordering for high-value shortlist formation while estimating hardware cost accurately enough to compare support points before expensive gem5 evaluation.

%%%%%%%%%%%%%%%%%%%%%%%%%%%%%%%%%%%%%%%%
%%%%%%%%%%%%%%%%%%%%%%%%%%%%%%%%%%%%%%%%

%%%%%%%%%%%%%%%%%%%%%%%%%%%%%%%%%%%%%%%%
%%%%%%%%%%%%%%%%%%%%%%%%%%%%%%%%%%%%%%%%

%%%%%%%%%%%%%%%%%%%%%%%%%%%%%%%%%%%%%%%%%%%%%%%%%%%%%%%%%%%%%%%%%%%%%%%%%%%%%%%%%%%%%%%%%%%%%%%%%%%%%%%%%%%%%%%%%%%%%%%%%%%%%%%
\subsection{Primitive-Side Efficiency Landscapes}
\label{subsec:eval_kernel}

\begin{figure}[t]
\centering
\includegraphics[width=\linewidth]{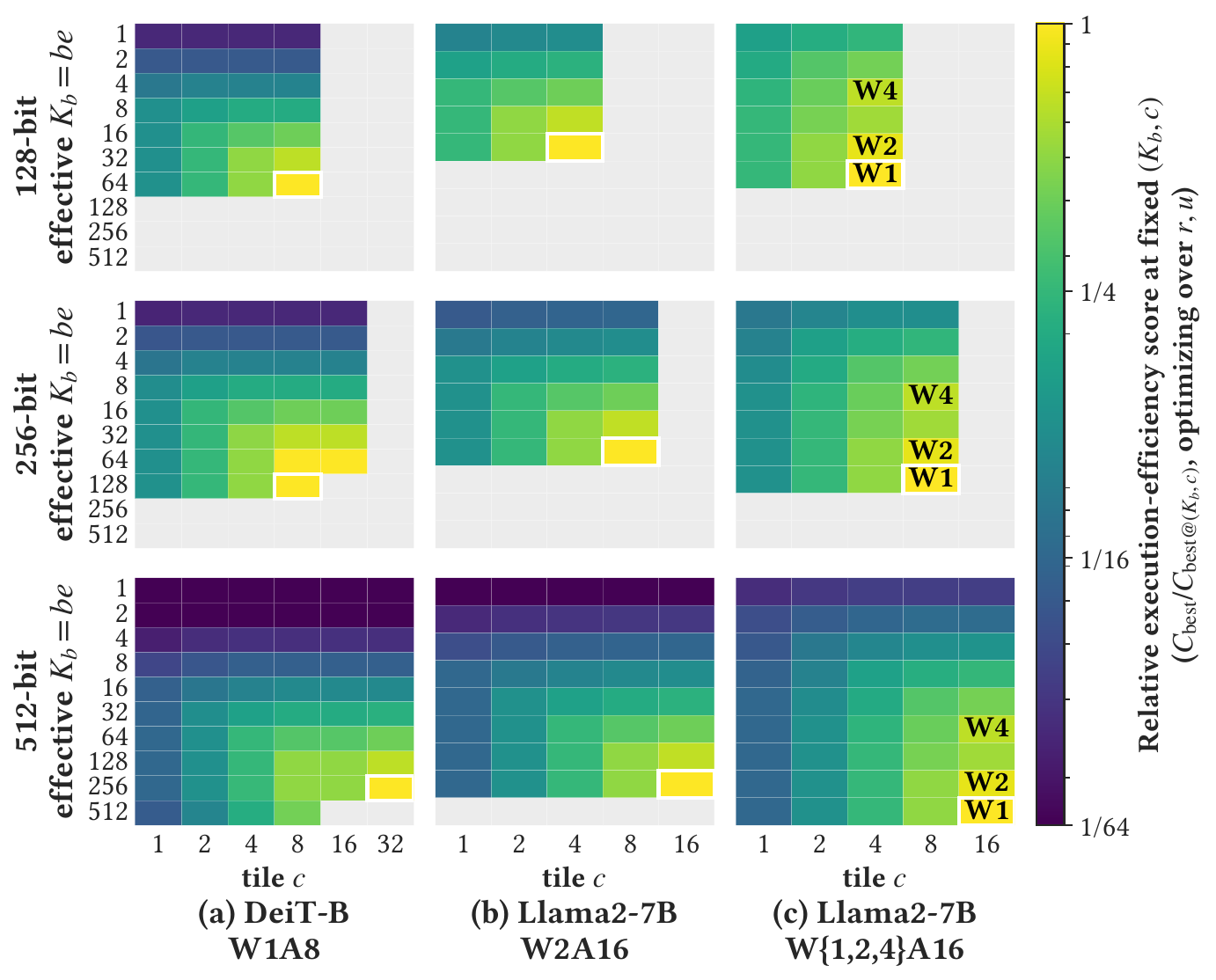}
\caption{Primitive-side analytical execution efficiency landscapes across workloads and targets. For each workload-target pair, the heatmap color at each $(K_b,c)$ reports the best relative analytical score at fixed effective K-block size $K_b = b e$ and column tile $c$ after optimizing over the remaining kernel parameters $(r,u)$ among feasible candidates; highlighted cells mark the best $(K_b,c)$ value(s). Uniform DeiT-B W1A8 and uniform Llama2-7B W2A16 each exhibit a compact target-scaled hotspot, whereas mixed-precision Llama2-7B W\{1,2,4\}A16 exhibits three regime-specific hotspots labeled W4/W2/W1.}
\label{fig:eval_kernel}
\end{figure}

We examine the primitive-side execution efficiency landscape under fixed constraints before hardware cost is introduced. \autoref{fig:eval_kernel} maps the relative analytical score over effective K-block size $K_b = b \cdot e$ and column tile $c$. For each workload-target pair, the color at each point reports the best relative analytical score attainable at fixed $(K_b,c)$ after optimizing over the remaining kernel parameters $(r,u)$ among feasible candidates. Equivalently, the plotted value is the normalized inverse execution cost, $C_{\mathrm{best}} / C_{\mathrm{best}@(K_b,c)}$, where $C_{\mathrm{best}@(K_b,c)}$ is the best modeled execution cost at that fixed primitive shape. The highlighted cells thus indicate feasible execution-favorable operating points rather than unconstrained primitive optima.

Two patterns emerge. First, the two uniform workloads, DeiT-B W1A8 and Llama2-7B W2A16, each exhibit a single dominant hotspot once the target width is fixed. As SIMD support widens, this hotspot shifts toward larger $K_b$ and $c$, indicating that the high-performance region is compact and largely target-scaled.

Second, the mixed-precision Llama2-7B W\{1,2,4\}A16 workload does not collapse to one hotspot. Instead, it exhibits three separated dominant regions aligned with the active quantization regimes, labeled W4, W2, and W1 in \autoref{fig:eval_kernel}. Across the 128/256/512-bit targets, these regime-specific hotspots move systematically toward larger $K_b$ and $c$. Thus, the mixed-precision workload is shaped not by arbitrary layer-to-layer noise, but by the superposition of several regime-specific primitive pressures within one model.

Selector overhead is introduced only when these primitive choices are aggregated into model-level support points. Thus, \autoref{fig:eval_kernel} isolates the primitive-side demand pattern in the execution space: the high-performance region is compact for the uniform workloads but split across several regime-specific hotspots for mixed-precision Llama2. We next ask whether this richer primitive-side structure is large enough to change the model-level support decision once hardware cost is included.

%%%%%%%%%%%%%%%%%%%%%%%%%%%%%%%%%%%%%%%%
%%%%%%%%%%%%%%%%%%%%%%%%%%%%%%%%%%%%%%%%

%%%%%%%%%%%%%%%%%%%%%%%%%%%%%%%%%%%%%%%%%%%%%%%%%%%%%%%%%%%%%%%%%%%%%%%%%%%%%%%%%%%%%%%%%%%%%%%%%%%%%%%%%%%%%%%%%%%%%%%%%%%%%%%
\subsection{End-to-End Comparative Evaluation}
\label{subsec:eval_e2e}

%%%%%%%%%%%%%%%%%%%%%%%%%%%%%%%%%%%%%%%%
\begin{figure}[t]
\centering
\includegraphics[width=\linewidth]{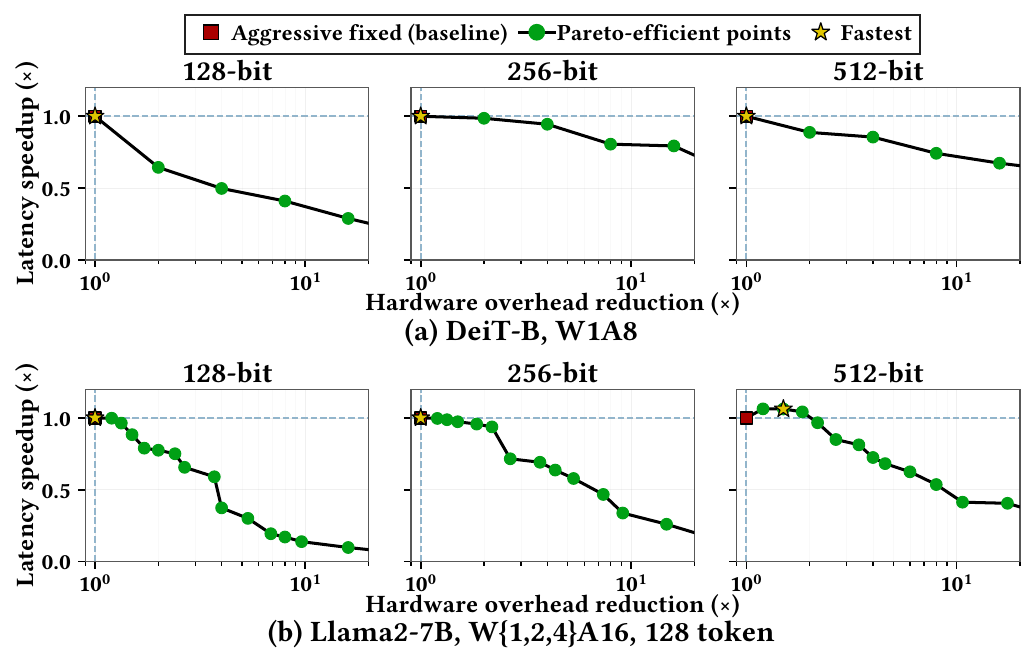}
\caption{Measured support-tradeoff frontier normalized to the aggressive fixed baseline. For each workload/target pair, the horizontal axis reports hardware overhead reduction relative to the aggressive fixed design, and the vertical axis reports latency speedup relative to that same baseline. DeiT-B stays close to the aggressive fixed point, whereas mixed-precision Llama2 exhibits a richer frontier, especially on the 512-bit target.}
\label{fig:graph_tradeoff}
\end{figure}
%%%%%%%%%%%%%%%%%%%%%%%%%%%%%%%%%%%%%%%%

\begin{figure}[t]
\centering
\includegraphics[width=\linewidth]{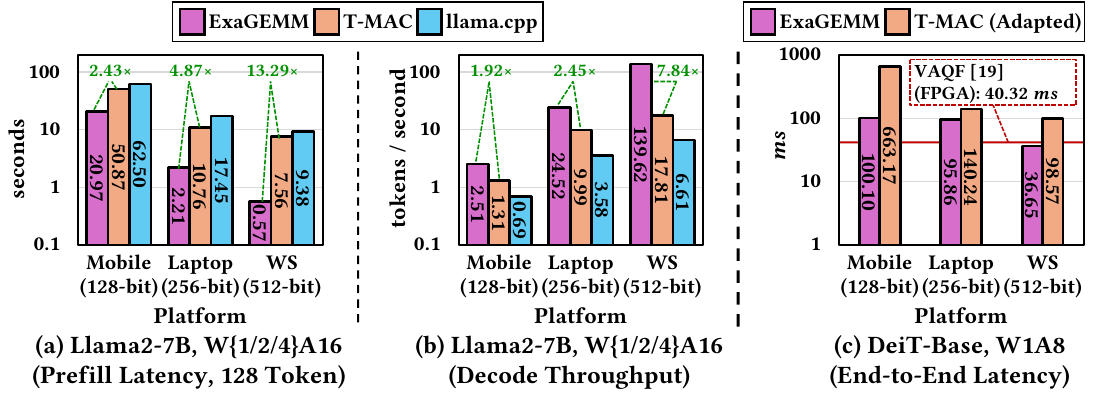}
\caption{End-to-end performance of \design{} against external baselines across the 128/256/512-bit targets. (a) Llama2-7B W\{1,2,4\}A16 prefill latency at 128 tokens. (b) Llama2-7B W\{1,2,4\}A16 decode throughput. (c) DeiT-B W1A8 end-to-end latency. AVX512-specific adaptations are applied to T-MAC and \texttt{llama.cpp}. \design{} outperforms the software baselines, with much larger widening gains on Llama than on DeiT-B.}
\label{fig:graph_comp}
\end{figure}

Finally, we evaluate whether the support points selected by \design{} remain attractive after gem5 materialization. This closes the design loop of \design{} by asking which primitive-side execution-favorable regions remain worth implementing once hardware overhead is included. \autoref{fig:graph_tradeoff} compares measured frontier points against an \emph{aggressive fixed} baseline, i.e., the feasible single-primitive design with the highest normalized support cost on each target, while \autoref{fig:graph_comp} reports end-to-end performance against external baselines.

\noindent\textbf{Frontier value depends on workload.}
In \autoref{fig:graph_tradeoff}, the aggressive fixed point for DeiT-B W1A8 lies on the measured frontier across all three targets, indicating that a single aggressive primitive is already a competitive support choice for the uniform workload. In contrast, Llama2-7B W\{1,2,4\}A16 exhibits a much richer frontier; on the 512-bit target, a measured frontier point is simultaneously faster and lower-cost than the aggressive fixed baseline. Thus, exploration still recovers the aggressive fixed design when it is Pareto-efficient, but the value of workload-aware frontier selection is much larger for mixed-precision LLM workloads than for DeiT-B.

%%%%%%%%%%%%%%%%%%%%%%%%%%%%%%%%%%%%%%%%
%%%%%%%%%%%%%%%%%%%%%%%%%%%%%%%%%%%%%%%%

\noindent\textbf{Widening benefits LLMs much more than ViTs.}
This difference is reinforced by \autoref{fig:graph_comp}, comparing \design{} to the prior SIMD kernels \cite{tmac, llama_cpp}. For \textbf{Llama2-7B W\{1,2,4\}A16}, \design{} reduces \textbf{prefill latency} by \textbf{2.43--13.29$\times$} and improves \textbf{decode throughput} by \textbf{1.92--7.84$\times$}. For \textbf{DeiT-B W1A8}, \design{} reduces \textbf{latency} by \textbf{1.46--6.62$\times$} over T-MAC. Although T-MAC was originally proposed for LLMs, its GEMM kernel can be adapted to low-bit ViT variants.

The widening payoff is similarly asymmetric. Moving from 256-bit to 512-bit support reduces Llama prefill latency by \textbf{3.89$\times$} and increases decode throughput by \textbf{5.69$\times$}, whereas DeiT-B improves by only \textbf{2.62$\times$}. DeiT-B also changes little from 128- to 256-bit, while Llama already improves sharply over that range. This is the end-to-end counterpart of \autoref{fig:eval_kernel}: DeiT-B is comparatively uniform, whereas mixed-precision Llama exposes a broader set of profitable primitive shapes and therefore benefits more from wider SIMD support and workload-aware selection.

Overall, \autoref{fig:graph_tradeoff} and \autoref{fig:graph_comp} show that a single aggressive primitive can be nearly sufficient for the uniform DeiT-B workload studied here, whereas for mixed-precision LLM workloads, workload-aware frontier selection changes the support decision itself and translates into substantial end-to-end gains.

%% file: Tables/system_cfg.tex
\begin{table}[t]
\centering
\caption{Representative target vector-CPU configurations used in gem5.}
\vspace{-4mm}
\label{tab:system_cfg}
\resizebox{0.9\linewidth}{!}{%
\begin{tabular}{c|ccc}
\toprule
\textbf{Attribute}  & \textbf{Workstation}  & \textbf{Laptop}   & \textbf{Mobile} \\
\midrule\midrule
CPU model           
& Ryzen 9 9950X
& Ryzen 7 6800U
& Cortex-A78 \\

Simulation mode
&
\multicolumn{3}{c}{DerivO3CPU}
\\

ISA                 
& x86-64
& x86-64
& ARMv8-A \\

Cores               
& 16
& 8
& 4 \\

Frequency           
& 5.7\,GHz
& 4.7\,GHz
& 3.0\,GHz \\

L1 I / D            
& 32\,KB / 48\,KB
& 32\,KB / 32\,KB
& 64\,KB / 64\,KB \\

L2 cache            
& 1\,MB/core
& 512\,KB/core
& 512\,KB/core \\

L3 cache            
& 64\,MB shared
& 16\,MB shared
& 4\,MB shared \\

DRAM                
& DDR5-6400
& DDR5-4800
& LPDDR5-6400 \\

VR param.s $l$ / $v$        
& 512\,b / 32
& 256\,b / 16
& 128\,b / 16 \\
\bottomrule
\end{tabular}}
\vspace{-4mm}
\end{table}

%% file: Chapters/6_Conclusion.tex
\vspace{-1mm}
\section{Conclusion}
\design{} formulates lightweight ISA support for low-bit CPU GEMM as a pre-silicon support-selection problem. By reusing existing SIMD datapaths for table generation and accumulation, it reduces new hardware to an in-register select network and treats hardware overhead as an explicit design constraint. \design{} prunes large candidate spaces before gem5, identifies compact non-dominated support frontiers, and generates implementation artifacts for system-level validation. The key result is workload-dependent: a single aggressive primitive is nearly sufficient for uniform DeiT-B, whereas mixed-precision Llama2 benefits substantially from workload-aware support selection. Across x86 and ARM targets, ExaGEMM delivers substantial gains over software-only and fixed-support baselines, providing a practical path to workload-aware low-bit CPU support across diverse quantization regimes.